
\input phyzzx  
\catcode`@=11 
\def\space@ver#1{\let\@sf=\empty \ifmmode #1\else \ifhmode
   \edef\@sf{\spacefactor=\the\spacefactor}\unskip${}#1$\relax\fi\fi}
\def\attach#1{\space@ver{\strut^{\mkern 2mu #1} }\@sf\ }
\newtoks\foottokens
\newbox\leftpage \newdimen\fullhsize \newdimen\hstitle \newdimen\hsbody
\newif\ifreduce  \reducefalse
\def\almostshipout#1{\if L\lr \count2=1
      \global\setbox\leftpage=#1 \global\let\lr=R
  \else \count2=2
    \shipout\vbox{\special{dvitops: landscape}
      \hbox to\fullhsize{\box\leftpage\hfil#1}} \global\let\lr=L\fi}
\def\smallsize{\relax
\font\eightrm=cmr8 \font\eightbf=cmbx8 \font\eighti=cmmi8
\font\eightsy=cmsy8 \font\eightsl=cmsl8 \font\eightit=cmti8
\font\eightt=cmtt8
\def\eightpoint{\relax
\textfont0=\eightrm  \scriptfont0=\sixrm
\scriptscriptfont0=\sixrm
\def\rm{\fam0 \eightrm \f@ntkey=0}\relax
\textfont1=\eighti  \scriptfont1=\sixi
\scriptscriptfont1=\sixi
\def\oldstyle{\fam1 \eighti \f@ntkey=1}\relax
\textfont2=\eightsy  \scriptfont2=\sixsy
\scriptscriptfont2=\sixsy
\textfont3=\tenex  \scriptfont3=\tenex
\scriptscriptfont3=\tenex
\def\it{\fam\itfam \eightit \f@ntkey=4 }\textfont\itfam=\eightit
\def\sl{\fam\slfam \eightsl \f@ntkey=5 }\textfont\slfam=\eightsl
\def\bf{\fam\bffam \eightbf \f@ntkey=6 }\textfont\bffam=\eightbf
\scriptfont\bffam=\sixbf   \scriptscriptfont\bffam=\sixbf
\def\tt{\fam\ttfam \eightt \f@ntkey=7 }
\def\caps{\fam\cpfam \tencp \f@ntkey=8 }\textfont\cpfam=\tencp
\setbox\strutbox=\hbox{\vrule height 7.35pt depth 3.02pt width\z@}
\samef@nt}
\def\Eightpoint{\eightpoint \relax
  \ifsingl@\subspaces@t2:5;\else\subspaces@t3:5;\fi
  \ifdoubl@ \multiply\baselineskip by 5
            \divide\baselineskip by 4\fi }
\parindent=16.67pt
\itemsize=25pt
\thinmuskip=2.5mu
\medmuskip=3.33mu plus 1.67mu minus 3.33mu
\thickmuskip=4.17mu plus 4.17mu
\def\thinspace{\kern .13889em }
\def\negthinspace{\kern-.13889em }
\def\enspace{\kern.416667em }
\def\enskip{\hskip.416667em\relax}
\def\quad{\hskip.83333em\relax}
\def\qquad{\hskip1.66667em\relax}
\def\crr{\cropen{8.3333pt}}
\foottokens={\Eightpoint\singlespace}
\def\papersize{\SIZE\OFFSET\skip\footins=\bigskipamount}
\def\SIZE{\hsize=11.8truecm\vsize=17.5truecm}
\def\OFFSET{\voffset=-1.3truecm\hoffset=  .14truecm}
\message{STANDARD CERN-PREPRINT FORMAT}
\def\attach##1{\space@ver{\strut^{\mkern 1.6667mu ##1} }\@sf\ }
\def\PH@SR@V{\doubl@true\baselineskip=20.08pt plus .1667pt minus .0833pt
             \parskip = 2.5pt plus 1.6667pt minus .8333pt }
\def\author##1{\vskip\frontpageskip\titlestyle{\tencp ##1}\nobreak}
\def\address##1{\par\kern 4.16667pt\titlestyle{\tenpoint\it ##1}}
\def\andaddress{\par\kern 4.16667pt \centerline{\sl and} \address}
\def\abstract{\vskip\frontpageskip\centerline{\twelvebf Astract}
              \vskip\headskip }
\def\cases##1{\left\{\,\vcenter{\Tenpoint\m@th
    \ialign{$####\hfil$&\quad####\hfil\crcr##1\crcr}}\right.}
\def\matrix##1{\,\vcenter{\Tenpoint\m@th
    \ialign{\hfil$####$\hfil&&\quad\hfil$####$\hfil\crcr
      \mathstrut\crcr\noalign{\kern-\baselineskip}
     ##1\crcr\mathstrut\crcr\noalign{\kern-\baselineskip}}}\,}
\Tenpoint
}
\def\Smallsize{\smallsize\reducetrue
\let\lr=L
\hstitle=8truein\hsbody=4.75truein\fullhsize=24.6truecm\hsize=\hsbody
\output={
  \almostshipout{\leftline{\vbox{\makeheadline
  \pagebody\makefootline}}}\advancepageno
     }
\special{dvitops: landscape}
\def\makeheadline{
\iffrontpage\line{\the\headline}
             \else\vskip .0truecm\line{\the\headline}\vskip .5truecm \fi}
\def\makefootline{\iffrontpage\vskip  0.truecm\line{\the\footline}
               \vskip -.15truecm\line{\the\date\hfil}
              \else\line{\the\footline}\fi}
\paperheadline={
\iffrontpage\hfil
               \else
               \tenrm\hss $-$\ \folio\ $-$\hss\fi    }
\paperstyle}
%
%
%
%
%
%
%
%
%
\newcount\referencecount     \referencecount=0
\newif\ifreferenceopen       \newwrite\referencewrite
\newtoks\rw@toks
\def\NPrefmark#1{\attach{\scriptscriptstyle [ #1 ] }}
\let\PRrefmark=\attach
\def\refmark#1{\relax\ifPhysRev\PRrefmark{#1}\else\NPrefmark{#1}\fi}
\def\refend{\refmark{\number\referencecount}}
\newcount\lastrefsbegincount \lastrefsbegincount=0
\def\refsend{\refmark{\count255=\referencecount
   \advance\count255 by-\lastrefsbegincount
   \ifcase\count255 \number\referencecount
   \or \number\lastrefsbegincount,\number\referencecount
   \else \number\lastrefsbegincount-\number\referencecount \fi}}
\def\refch@ck{\chardef\rw@write=\referencewrite
   \ifreferenceopen \else \referenceopentrue
   \immediate\openout\referencewrite=referenc.texauxil \fi}
%
{\catcode`\^^M=\active 
  \gdef\obeyendofline{\catcode`\^^M\active \let^^M\ }}%
%
{\catcode`\^^M=\active 
  \gdef\ignoreendofline{\catcode`\^^M=5}}
{\obeyendofline\gdef\rw@start#1{\def\t@st{#1} \ifx\t@st\blankend%
\endgroup \@sf \relax \else \ifx\t@st\bl@nkend \endgroup \@sf \relax%
\else \rw@begin#1
\backtotext
\fi \fi } }
{\obeyendofline\gdef\rw@begin#1
{\def\n@xt{#1}\rw@toks={#1}\relax%
\rw@next}}
\def\blankend{}
{\obeylines\gdef\bl@nkend{
}}
\newif\iffirstrefline  \firstreflinetrue
\def\rwr@teswitch{\ifx\n@xt\blankend \let\n@xt=\rw@begin %
 \else\iffirstrefline \global\firstreflinefalse%
\immediate\write\rw@write{\noexpand\obeyendofline \the\rw@toks}%
\let\n@xt=\rw@begin%
      \else\ifx\n@xt\rw@@d \def\n@xt{\immediate\write\rw@write{%
        \noexpand\ignoreendofline}\endgroup \@sf}%
             \else \immediate\write\rw@write{\the\rw@toks}%
             \let\n@xt=\rw@begin\fi\fi \fi}
\def\rw@next{\rwr@teswitch\n@xt}
\def\rw@@d{\backtotext} \let\rw@end=\relax
\let\backtotext=\relax

\newdimen\refindent     \refindent=30pt
\def\refitem#1{\par \hangafter=0 \hangindent=\refindent \Textindent{#1}}
\def\REFNUM#1{\space@ver{}\refch@ck \firstreflinetrue%
 \global\advance\referencecount by 1 \xdef#1{\the\referencecount}}
\def\refnum#1{\space@ver{}\refch@ck \firstreflinetrue%
 \global\advance\referencecount by 1 \xdef#1{\the\referencecount}\refend}

\def\REF#1{\REFNUM#1%
 \immediate\write\referencewrite{%
 \noexpand\refitem{#1.}}%
\begingroup\obeyendofline\rw@start}
\def\ref{\refnum\?%
 \immediate\write\referencewrite{\noexpand\refitem{\?.}}%
\begingroup\obeyendofline\rw@start}
\def\Ref#1{\refnum#1%
 \immediate\write\referencewrite{\noexpand\refitem{#1.}}%
\begingroup\obeyendofline\rw@start}
\def\REFS#1{\REFNUM#1\global\lastrefsbegincount=\referencecount
\immediate\write\referencewrite{\noexpand\refitem{#1.}}%
\begingroup\obeyendofline\rw@start}
\newif\ifmref  
\newif\iffref  
\def\xrefsend{\xrefmark{\count255=\referencecount
\advance\count255 by-\lastrefsbegincount
\ifcase\count255 \number\referencecount
\or \number\lastrefsbegincount,\number\referencecount
\else \number\lastrefsbegincount-\number\referencecount \fi}}
\def\xrefsdub{\xrefmark{\count255=\referencecount
\advance\count255 by-\lastrefsbegincount
\ifcase\count255 \number\referencecount
\or \number\lastrefsbegincount,\number\referencecount
\else \number\lastrefsbegincount,\number\referencecount \fi}}
\def\xREFNUM#1{\space@ver{}\refch@ck\firstreflinetrue%
\global\advance\referencecount by 1
\xdef#1{\xrefend}}
\def\xrefend{\xrefmark{\number\referencecount}}
\def\xrefmark#1{[{#1}]}
\def\xRef#1{\xREFNUM#1\immediate\write\referencewrite%
{\noexpand\refitem{#1 }}\begingroup\obeyendofline\rw@start}%
\def\xREFS#1{\xREFNUM#1\global\lastrefsbegincount=\referencecount%
\immediate\write\referencewrite{\noexpand\refitem{#1 }}%
\begingroup\obeyendofline\rw@start}
\def\rrr#1#2{\relax\ifmref{\iffref\xREFS#1{#2}%
\else\xRef#1{#2}\fi}\else\xRef#1{#2}\xrefend\fi}
\def\multref#1#2{\mreftrue\freftrue{#1}%
\freffalse{#2}\mreffalse\xrefsend}
\def\doubref#1#2{\mreftrue\freftrue{#1}%
\freffalse{#2}\mreffalse\xrefsdub}
\referencecount=0
\def\refbreak{\hfil\penalty200\hfilneg}
\def\paperstyle{\papers}
\paperstyle   
%
%
%
\def\slacpub{\afterassignment\slacp@b\toks@}
\def\slacp@b{\edef\n@xt{\Pubnum={NIKHEF--H/\the\toks@}}\n@xt}
\let\pubnum=\slacpub
\expandafter\ifx\csname eightrm\endcsname\relax
    \let\eightrm=\ninerm \let\eightbf=\ninebf \fi

\font\seventeencp=cmcsc10 scaled\magstep3

\newif\ifCONF \CONFfalse
\newif\ifBREAK \BREAKfalse
\newif\ifsectionskip \sectionskiptrue

%
%
%
%
\def\NuclPhysProc{
\let\lr=L
\hstitle=8truein\hsbody=4.75truein\fullhsize=21.5truecm\hsize=\hsbody
\hstitle=8truein\hsbody=4.75truein\fullhsize=20.7truecm\hsize=\hsbody
\output={
  \almostshipout{\leftline{\vbox{\makeheadline
  \pagebody\makefootline}}}\advancepageno
     }
\def\papersize{\SIZE\OFFSET\skip\footins=\bigskipamount}
\def\SIZE{\hsize=10.0truecm\vsize=27.0truecm}
\def\OFFSET{\voffset=-1.4truecm\hoffset=-2.40truecm}
\message{NUCLEAR PHYSICS PROCEEDINGS FORMAT}
\def\makeheadline{
\iffrontpage\line{\the\headline}
             \else\vskip .0truecm\line{\the\headline}\vskip .5truecm \fi}
\def\makefootline{\iffrontpage\vskip  0.truecm\line{\the\footline}
               \vskip -.15truecm\line{\the\date\hfil}
              \else\line{\the\footline}\fi}
\paperheadline={\hfil}
\paperstyle}
%
%
%
%

%
%
%
%

%
%
%
%
\def\ReprintVolume{\smallsize
\def\papersize{\hsize=18.0truecm\vsize=25.1truecm\voffset -.73truecm
    \hoffset -.65truecm\skip\footins=\bigskipamount
    \normaldisplayskip= 20pt plus 5pt minus 10pt}
\message{REPRINT VOLUME FORMAT}
\paperstyle\baselineskip=.425truecm\parskip=0truecm
\def\makeheadline{
\iffrontpage\line{\the\headline}
             \else\vskip .0truecm\line{\the\headline}\vskip .5truecm \fi}
\def\makefootline{\iffrontpage\vskip  0.truecm\line{\the\footline}
               \vskip -.15truecm\line{\the\date\hfil}
              \else\line{\the\footline}\fi}
\paperheadline={
\iffrontpage\hfil
               \else
               \tenrm\hss $-$\ \folio\ $-$\hss\fi    }
\def\sectionfont{\bf}    }
%
%
%
%
\def\SIZE{\hsize=15.73truecm\vsize=23.11truecm}
\def\OFFSET{\voffset=0.0truecm\hoffset=0.truecm}
\message{DEFAULT FORMAT}
\def\papersize{\SIZE\OFFSET\skip\footins=\bigskipamount
\normaldisplayskip= 35pt plus 3pt minus 7pt}
\Pubnum={\rm NIKHEF--H/\the\pubnum }
\def\title#1{\vskip\frontpageskip\vskip .50truein
     \titlestyle{\seventeencp #1} \vskip\headskip\vskip\frontpageskip
     \vskip .2truein}
\def\author#1{\vskip .27truein\titlestyle{#1}\nobreak}

\def\p@bblock{\begingroup \tabskip=\hsize minus \hsize
   \baselineskip=1.5\ht\strutbox \topspace+6\baselineskip
   \halign to\hsize{\strut ##\hfil\tabskip=0pt\crcr
  \the \Pubnum\cr}\endgroup}
\def\makefootline{\iffrontpage\vskip .27truein\line{\the\footline}
                 \vskip -.1truein
              \else\line{\the\footline}\fi}
\paperfootline={\iffrontpage\message{FOOTLINE}
\hfil\else\hfil\fi}
\def\abstract{\vskip\frontpageskip\centerline{\twelvebf Astract}
              \vskip\headskip }
\paperheadline={
\iffrontpage\hfil
               \else
               \twelverm\hss $-$\ \folio\ $-$\hss\fi}
%
%
\def\nup#1({\refbreak\ Nucl.\ Phys.\ $\underline {B#1}$\ (}
\def\plt#1({\refbreak\ Phys.\ Lett.\ $\underline  {#1}$\ (}
\def\cmp#1({\refbreak\ Commun.\ Math.\ Phys.\ $\underline  {#1}$\ (}
\def\prp#1({\refbreak\ Physics\ Reports\ $\underline  {#1}$\ (}
\def\prl#1({\refbreak\ Phys.\ Rev.\ Lett.\ $\underline  {#1}$\ (}
\def\prv#1({\refbreak\ Phys.\ Rev. $\underline  {D#1}$\ (}
\def\und#1({            $\underline  {#1}$\ (}
%
%

\def\rB{\hfil\penalty1000\hfilneg}
%
%
\hyphenation{sym-met-ric anti-sym-me-tric re-pa-ra-me-tri-za-tion
Lo-rentz-ian a-no-ma-ly di-men-sio-nal two-di-men-sio-nal}
%
%
%
%

\def\coeff#1#2{{\textstyle { #1 \over #2}}\displaystyle}
\def\boxit#1{\vbox{\hrule\hbox{\vrule\kern3pt
\vbox{\kern3pt#1\kern3pt}\kern3pt\vrule}\hrule}}
\message{ by V.K, W.L and A.S}
\catcode`@=12
\paperstyle
\paperstyle
\def\KLST{\rrr\KLST{H.~Kawai, D.~Lewellen, J.A.~Schwartz
and S.-H.H.~Tye \nup 299 (1988) 431.}}
\def\GrS {\rrr\GrS {M.~Green and J.~Schwarz, \plt 149B (1984) 117.}}
\def\KLTA{\rrr\KLTA{
H.~Kawai, D.~Lewellen and S.~Tye,
\nup288 (1987) 1.}}
\def\ABK{\rrr\ABK{
I.~Antoniadis, C.~Bachas and C.~Kounnas,
\nup289 (1987) 87.}}
\def\BDGA{\rrr\BDGA{
R.~Bluhm, L.~Dolan and P.~Goddard,
\nup309 (1988) 330.}}
\def\ThM {\rrr\ThM {J.~Thierry-Mieg, \plt 171B (1986) 163.}}
\def\SchM{\rrr\SchM{A.N.~Schellekens, {\it Meromorphic c=24
Conformal Field Theories}, CERN preprint, CERN-TH.6478/92,
to appear in Commun. Math. Phys.}}
\def\Ven {\rrr\Ven {B.B Venkov, {\it The Classification of Integral
Even Unimodular Quadratic Forms}, Proc. of the Steklov
Institute of Mathematics 4 (1980) 63.}}
\def\Mon {\rrr\Mon {P.~Montague, {\it Discussion of Self-Dual c=24
Conformal Field Theories}, Cambridge preprint, hep-th/920567.}}
\def\FHPV{\rrr\FHPV{
P.~Forgacs, L.~Palla,  Z.~Horvath and P.~Vecserny\'es,
\nup308 (1988) 477.}}
\def\DrFu{\rrr\DrFu{J.~Fuchs and P. van Driel, Lett. Math. Phys. 23
 (1991) 11.}}
\def\Lee {\rrr\Lee {J.~Leech, Canadian Journal of Mathematics
\und 19 (1967) 251.}}
\def\Sie {\rrr\Sie {C.~Siegel, Ann.~Math.~\und  36 (1935) 527.}}
\def\CoS{\rrr\CoS{
J.~Conway and M.~Sloane, Journal of Number Theory \und 15 (1982) 83; \rB
Europ.~J.~Combinatorics 3 (1982) 219.}}
\def\FQS{\rrr\FQS{D.~Friedan, Z.~Qiu and S.~Shenker,
\prl52  (1984) 1575.}}
\def\ScW {\rrr\ScW {A.N.~Schellekens
and N.P.~Warner, \plt B177 (1986) 317;
\plt B181 (1986) 339;     \nup287 (1987) 317.}}
\def\VGM{\rrr\VGM{It was pointed out by Vafa, Ginsparg and Moore that
the conclusions of \ScW\ can be generalized immediately to arbitrary
Kac-Moody algebras using the results of Kac and Peterson on the
transformation properties of Kac-Moody characters; C. Vafa,
private communication; P. Ginsparg
{\it Informal String Lectures},
 in the Proceedings of the U.K.~Institute for Theoretical
 High Energy Physics, Cambridge, 16 Aug.~- 5 Sept.~1987.}}
\def\FGK {\rrr\FGK {G.~Felder, K.~Gawedzki and A.~Kupainen,
\cmp 117 (1988) 127.}}
\def\DiH{\rrr\DiH{
L.~Dixon and J.~Harvey, \nup274 (1986) 93.}}
\def\LLSA{\rrr\LLSA{
W.~Lerche, D.~L\"ust and A.N.~Schellekens, \plt B181 (1986) 71;
Erratum, \plt B184 (1987) 419.}}
\def\SchH{\rrr\SchH{
A.N.~Schellekens,
\plt B277 (1992) 277.}}
\def\KaPe{\rrr\KaPe{
V.G.~Kac and D.H.~Peterson, Adv.~in Math.~53 (1984) 125.}}
\def\Nie{\rrr\Nie{
H.~Niemeier, Journal of Number Theory \und 5 (1973) 142.}}
\def\GKO{\rrr\GKO{
P.~Goddard, A.~Kent and D.~Olive,
\plt152B (1985) 88;\rB \cmp103 (1986) 105.}}
\def\CoSu{\rrr\CoSu{F.~Bais and
P.~Bouwknegt,
\nup279 (1987) 561; \rB A.N.~Schellekens and N.P.~Warner,
\prv34 (1986) 3092.}}
\def\BPZ {\rrr\BPZ {A.~Belavin, A.~Polyakov and A.~Zamolodchikov,
 \nup241 (1984) 333.}}
\def\AGMV{\rrr\AGMV{L.~Alvarez-Gaum\'e, P.~Ginsparg, G.~Moore and
                    C.~Vafa, \plt B171 (1986) 155.}}
\def\KLTE{\rrr\KLTE{H.~Kawai, D.~Lewellen and S.~Tye,
\prv34 (1986) 3794.}}
\def\ScYc{\rrr\ScYc{A.N.~Schellekens and S.~Yankielowicz,
\plt B226 (1989) 285.}}
\def\BB  {\rrr\BB  {L.~Dixon, P.~Ginsparg and J.~Harvey,
\cmp119 (1988) 221.}}
\def\God {\rrr\God {P.~Goddard, {\it Meromorphic Conformal Field Theory},
preprint DAMTP-89-01.}}
\def\EVeA{\rrr\EVeA{E.~Verlinde,
\nup300 (1988) 360.}}
\def\ZamA{\rrr\ZamA{A.~Zamolodchikov, Theor.\ Math.\ Phys.\
\und 65 (1986) 1205.}}
\def\LSW {\rrr\LSW {W.~Lerche, A.~Schellekens and N.~Warner,
\plt B214 (1988) 41.}}
\def\AhWa{\rrr\AhWa{C.~Ahn and M.~Walton,\plt B223 (1989) 343.}}
\def\Gin {\rrr\Gin {P.~Ginsparg, {\it Informal String Lectures},
 in the Proceedings of the U.K.~Institute for Theoretical
 High Energy Physics, Cambridge, 16 Aug.~- 5 Sept.~1987, Harvard preprint
 HUTP-87/A077 (1987).}}
\def\DVVV{\rrr\DVVV{R.~Dijkgraaf, C.~Vafa, E.~Verlinde and
H.~Verlinde,
\cmp 123 (1989) 16.}}
\def\Kir {\rrr\Kir {E.~Kiritsis, \nup324 (1989) 475.}}
%
\def\ScYb{\rrr\ScYb{
A.N.~Schellekens and S.~Yankielowicz,
\plt B227 (1989) 387.}}
\def\ScYA{\rrr\ScYA{
A.N.~Schellekens and S.~Yankielowicz,
\nup 327 (1989) 673; \rB
\plt B227 (1989) 387.}}
\def\CIZ {\rrr\CIZ {
A.~Cappelli, C.~Itzykson     and J.-B.~Zuber,
\nup280 (1987)  445;\rB \cmp113 (1987) 1.}}
\def\Ber {\rrr\Ber {
D.~Bernard, \nup288 (1987) 628.}}
\def\ALZ {\rrr\ALZ {D.~Altschuler, J.~Lacki and P.~Zaugg, \plt B205
(1988) 281.}}
\def\Alig{\rrr\Alig{K.~Intriligator,
\nup 332 (1990) 541.}}
\def\Fuch{\rrr\Fuch{J.~Fuchs,
\cmp136 (1991) 345.}}
\def\BeBT{\rrr\BeBT {B.~Gato-Rivera and A.N.~Schellekens,
\cmp 145 (1992)  85.}}
\def\DGM {\rrr\DGM  {L.~Dolan, P.~Goddard and P.~Montague,
\plt B236 (1990) 165.}}
\def\Ver {\rrr\Ver  {D.~Verstegen, \nup 346 (1990) 349;
\cmp 137 (1991) 567.}}
\def\LFM {\rrr\LFM  {I.~Frenkel, J.~Lepowsky and J.~Meurman,
Proc. Natl. Acad. Sci. U.S.A. 81, 32566 (1984).}}
\def\KaWa{\rrr\KaWa{V.Kac and M. Wakimoto, Adv. Math. 70 (1988) 156.}}
\def\Walt{\rrr\Walt{M. Walton, \nup322 (1989) 775.}}
\def\ABI {\rrr\ABI {D. Altschuler, M. Bauer and C. Itzykson,
\cmp132 (1990) 349.}}
\def\Pat {\rrr\Pat {J. Math. Phys. 17 (1976) 1972;
S. Okubo and J. Patera, J. Math. Phys. 24 (1983)
2722; 25 (1984) 219. }}
\hfuzz= 6pt
\def\Zbf{{\bf Z}}

\def\half{\coeff12}

\def\N{{\cal N}}
\def\X{{\cal X}}
\def\ie{i.e.}
\def\eg{e.g.}

\pubnum={{93--08}}
\date{April 1993}
\pubtype{CRAP}
\titlepage
\message{TITLE}
\title{\fourteenbf Seventy Relatives of the Monster Module\foot{Presented
at the
$16^{\rm th}$ Johns Hopkins workshop on Current Problems in Particle
Theory, G\"oteborg,
June 8-10 1992, the
International
Conference on  Modern  Problems  in  Quantum  Field  Theory,   Strings
and Quantum Gravity, Kiev, June 8-17 1992, the
$4^{\rm th}$ Hellenic Summer School on Elementary Particle Physics,
Corfu,September 1992, and
the Workshop on
Supergravity and Superstrings, Erice, December 5-12 1992.}}
\author{A. N. Schellekens}
\line{\hfil \tenpoint\it NIKHEF-H, P.O. Bos 41882, 1009 DB Amsterdam,
The Netherlands  \hfil}
\abstract
Recent work on the
classification
of conformal field theories with one
primary field (the identity operator) is reviewed. The classification of
such theories is an essential step in the program of classification of all
rational conformal field theories, but appears impossible in general.
The last managable case, central charge 24, is considered here. We found a
total of 71 such theories (which have not all been
constructed yet), including the monster module. The complete list of
modular invariant partition functions has already appeared elsewhere \SchM.
This paper contains
an easily readable account of the method, as well as a few examples and
some comments.
\vfill\endpage

\chapter{Introduction}

The simplest conformal field theories, from the point of view
of the modular group or the fusion rules, are those with just
one primary field with respect to some integer spin chiral algebra.
It is elementary to show that unitary conformal field theories
of this kind must have a central charge that is a multiple of 8. They
transform according to a one-dimensional representation of the
modular group with $S=1$ and $T$ a cubic root of unity. Furthermore,
if the central charge is a multiple of 24 the single character is
modular invariant by itself, and can be written as a polynomial in
the absolute modular invariant $j$,
$$ j= {1\over q } + 744 + 196884 q + 21493760 q^2 + \ldots   \ , $$
with a leading term $q^{-n}$ if the central charge $c=24n$; here
$q=e^{2\pi i \tau}$. Since the character ${\cal X}$ is modular invariant
by itself
one may consider,
instead of the usual ``diagonal'' CFT with partition function
${\cal X}{\cal X}^*$, a
purely chiral conformal field theory with partition function
${\cal X}$. Such a theory will be called
a {\it meromorphic} conformal
field theory, and denoted MCFT.

The classification of these theories is an essential part of
the programme of classification of rational conformal field theories,
initiated a few years ago. Indeed, one can argue
that the entire RCFT classification problem can be embedded in that
of classification of MCFT's, provided that one can
show that any RCFT has a {\it complement}. This is
a RCFT with the same
number of primary fields and complex conjugate $S$ and $T$ matrices.
(A complement can easily be constructed
for all WZW-models and for all coset theories without
field identification fixed points).
Then any diagonal RCFT with a
modular invariant $\sum_i {\cal X}_i({\cal X}_i)^*$
can be mapped to a meromorphic one with partition function
$\sum_i {\cal X}_i {\cal X}^C_i $, where `$C$' denotes the
complement.

In any case it is clear that the RCFT classification problem is not
solved as long as we cannot even classify the theories with just
one primary field.
This is bad news, since for $c\geq32$
the number of such theories grows so fast with the central
charge that listing them is simply impossible. Indeed, for $c=32$
the number of such theories is known to be larger that $8 \times 10^7$.
The problem looks
substantially
easier for $c\leq 24$, and with some
(though probably unfounded) optimism one may hope that
the information contained in the  $c\geq32$ theories will never really be
needed in practice.

The fact that enumeration is impossible for $c\geq32$ may dampen ones
enthusiasm for attempting a enumeration for $c\leq24$.
Nevertheless, there
are indications that the $c\leq24$ theories (and in particular those
with $c=24$) are of some intrinsic
interest. In physics, $c=24$ is special because of the bosonic string,
whose transverse dimension is 24; in mathematics the number 24 plays
a special role in many contexts, such as the theory of sphere
packings or the Monster group (the largest of the sporadic simple
finite groups), for which a meromorphic $c=24$ theory provides a
``natural'' $q$-graded representation, the ``monster module'' \LFM. One may
hope that a list of the CFT-relatives of the monster module places this
object in a new and interesting context.
These admittedly
rather vague motivations will probably turn out to be
the most important ones for
attempting to classify the meromorphic $c=24$ CFT's.
A somewhat more
practical motivation is that a listing of such theories has enabled us
to complete another classification problem, that of ten-dimensional
heterotic strings \SchH. Yet another unsolved problem about
which we have learned a few interesting new facts (without solving it,
though) is that of the classification of Kac-Moody modular invariants.
Several new non-diagonal invariants of simple Kac-Moody algebras were
found that are 'highly exceptional': they are not simple current
invariants or conformal embeddings, nor are they related to such
invariants by rank-level duality.

A large class of MCFT's can be constructed by taking $8n$
free bosons with momenta quantized on an even self-dual lattice. This
gives $1, 2$ and $24$ \doubref\Lee\Nie\
distinct theories for $c=8,16$ and $24$
respectively
(and more than $8\times 10^7$ for $c=32$).
This class can be enlarged
by a $\Zbf_2$ orbifold twist, using the symmetry that sends every
boson $X$ to $-X$ \doubref\BB\DGM.
This gives back the same $E_{8,1}$ theory
for $c=8$, and maps the two $c=16$ MCFT's $(E_{8,1})^2$ and
$D_{16,1}([0]+[s])$ to one another (the argument denotes the
conjugacy classes that appear). The result is more interesting when
this twist is applied to the Leech lattice and the 23 Niemeier lattices:
The former gives a new MCFT, the monster module, while from the
latter one gets other Niemeier lattices in 9 cases, and new
MCFT's in the 14 remaining cases \DGM.
Altogether this gives us thus
$1,2$ and 39 MCFT's for $c=8,16$ and $24$.

Clearly there are other orbifold twists one might consider, but
it becomes rather difficult to prove the consistency of the resulting
theories. More importantly, even an exhaustive classification of
all orbifolds of known theories is not sufficient to show that
the result is complete. The same is true for other kinds of
constructions. For example, one could study all tensor products of
Kac-Moody algebras with total central charge $8n$, and determine their
meromorphic modular invariants. Even though this is a finite problem,
there is no guarantee that the answer will be complete, since in
general only part of the central charge will be saturated by
(non-abelian) Kac-Moody algebras. As soon as one allows rational
$U(1)$ factor the problem is not finite anymore,
and it gets still worse
if one adds factors without spin-1 currents (e.g. coset theories).
In any case, it was already known for some time that the number
of MCFT's with $c=24$ is larger than the 39 mentioned so far: two
additional candidates were presented in \ScYc, one of which can
certainly be constructed explicitly.

While explicit constructions approach the set of solutions from below
it is possible in some cases to limit the set of solutions from above,
i.e. to derive necessary rather than sufficient conditions for the
existence of solutions. An example is the set of $c=8$ and $c=16$
solutions. Any such theory can be used to build a supersymmetric
heterotic string theory in 10 dimensions. It can be shown in general
(\ScW, see also \VGM)
that modular invariance of such a theory implies that all gauge
and gravitational anomalies of the resulting field theory must
factorize \`a la Green-Schwarz \GrS.
But all possibilities for such
anomaly cancellations are known \GrS\ThM,
and this immediately reduces the
$c=8$ and $c=16$ theories to $(E_{8,1})^2$ and $D_{16,1}$. There cannot
exist more such theories, and since both can be constructed
using self-dual lattices, there are no fewer either.

It turns out that a similar argument can be applied, with a
considerably larger effort, to the $c=24$ theories \SchM.
Beyond $c=24$ the nature of the problem changes drastically, and
these methods become useless, not just in practice but even in principle.
The basic idea is to write down a character valued partition function
for a given $c=24$ theory analogous to similar functions introduced
in \ScW\ for the chiral sector of heterotic strings. This function
generalizes the ordinary one-loop partition function
$$ P(q) = \sum_{n=1}^{\infty} d_n q^n \ , $$
by replacing the multiplicities $d_n$ by Chern-characters of the
representation at each level. Thus we get
$$ P(q,F)=\sum_{n=1}^{\infty} \Tr e^F q^n \ . $$
Here $F$ is some representation matrix of a semi-simple
Lie-algebra, in the representation of the $n^{\rm th}$  level.

To write down such a partition function we must have a Lie
algebra that organizes the levels according to its representations.
This happens if the theory has a set of spin-1 currents, which
necessarily close into a Kac-Moody algebra, plus possibly some
$U(1)$-currents \ZamA. Note that at this point we are certainly not
assuming that these algebras saturate the central charge.

In general, the Kac-Moody algebra
consists
of several simple factors, and the partition function can be expressed in
terms of the characters $\X^{\ell}_{i_{\ell}}$ of the ${\ell}^{\rm th}$
factor and an unknown function without spin-1 contributions:
$$ P(q,\vec F_1,\ldots,\vec F_L)
=\sum_{i_1,\ldots,i_L}
\X^1_{i_1}(q,\vec F_1) \ldots \X^L_{i_L}(q,\vec F_L)
                 \X_{i_1,\ldots i_L}(q) \ . $$
Here $\vec F_{\ell}$ denotes the decomposition of $F$ with
respect to a basis of Lie-algebra generators $J_0^a$ in each of the
simple factors: $F=\sum_a F^a J_0^a=\vec F \cdot \vec J_0$.
Now we wish to make use of the modular transformation properties
of the theory. For $c=8n$ $P$ transforms with $S=1$ and
$T=e^{-2 \pi i n /3}$. It is convenient to multiply $P$ with
$\eta(q)^{8n}$ to remove the phase in the $T$ transformation.
Then the function
$\hat P(q,0,\ldots,0)=[\eta(q)]^{8n} P(q,0,\ldots,0)$ transforms
as a modular function of weight $4n$.
Furthermore we know the transformation properties
of the Kac-Moody characters \KaPe\
$$ \eqalign {\tau \rightarrow \tau+1 \ &: \ \ \
  {\cal X}_i(\tau+1,\vec F)=e^{2\pi i (h_i - c/24)  }
   {\cal X}_i(\tau,\vec F) \cr
 \tau \rightarrow -{1\over \tau} \ &: \ \ \
  {\cal X}_i(-{1\over\tau}, {\vec F\over\tau})=
e^{-{i\over 8 \pi  \tau}
{k\over g} \Tr_{\rm adj} F^2 }S_{ij}
{\cal X}_j(\tau,\vec F)\ , \cr }\eqn\Trafo$$
where
$$\X_i(\tau,\vec F)=\Tr_i e^{\vec F \cdot \vec J_0}
                    e^{2\pi i \tau (L_0 -c/24)}\ , \eqn\Character$$
with the trace evaluated over the positive norm states of the
representation ``$i$''. In \Trafo\
$g$ is the dual Coxeter number of the Kac-Moody algebra, and we
have traded $q$ for $\tau$, with $q=e^{2\pi i \tau}$. The trace
in \Trafo\ is evaluated in the adjoint
representation\rlap.\foot{Conventions: $J_0^a$ is Hermitian,
$f_{abc}f_{abe}=2g\delta_{ce}$.
For
$U(1)$ factors the adjoint representation is not suitable, but
one can use any non-trivial
representation, provided that $k/g$ is replaced by some
normalization $N$. This will be implicitly assumed in the following.}

Using \Trafo\ and the fact that the $\hat P$ must be a
modular function for $\vec F=0$, we can derive how it must transform
when $\vec F\not=0$.
One finds
$$ \hat P\left({a\tau+b\over c\tau+d},{\vec F\over c\tau+d}\right) =
  \exp\left[{-ic\over 8 \pi (c\tau+d)} {\cal F}^2
           \right]\  (c\tau+d)^{4n}
                \hat P(\tau,\vec F)  \ ,\eqn\TrafoTwo  $$
where we have defined
$$ {\cal F}^2 = \sum_{\ell} {k_{\ell}\over g_{\ell}}
\Tr_{\rm adj} F_{\ell}^2\ . $$

To analyse the consequences of these transformation properties we need
the Eisenstein functions,  for convenience normalized as follows
$$\eqalign{  E_2(q)&=1-24\sum_{n=1}^{\infty}
             { n   q^n \over 1 -  q^n } \ , \cr
             E_4(q)&=1+240\sum_{n=1}^{\infty}
             { n^3 q^n \over 1 -  q^n } \ , \cr
             E_6(q)&=1-504\sum_{n=1}^{\infty}
             { n^5 q^n \over 1 -  q^n } \ , \cr} $$
The last two are entire modular functions of
weight 4 and 6 respectively, whereas $E_2$ has an anomalous
term in its modular transformation
$$ E_2\left({a\tau+b\over  c\tau+d}\right)=
   (c\tau+d)^2E_2(\tau) - {6i\over\pi} c (c\tau+d) \ . $$

The anomalous term in the $E_2$ transformation can be used to cancel
the exponential prefactor in \TrafoTwo. Indeed, if we define
$$\tilde P(q,\vec F) =e^{-{1/48}E_2(q){\cal F}^2} \hat P(q,\vec F) $$
we find
$$ \tilde P\left({a\tau+b\over c\tau+d},{\vec F\over c\tau+d}\right) =
(c\tau+d)^{4n}   \tilde P(\tau,\vec F)  \ .\eqn\TrafoThree  $$
Expanding $\tilde P$ in powers of $F$ one finds that the expansion
coefficients of terms of order $m$ must be modular functions of
weight $4n+m$. Furthermore they do not have poles at $\tau=i\infty$
because  we have taken out the required number of $\eta$-functions.
It will be necessary to {\it assume} that they do not have poles
elsewhere in the upper half-plane. This is automatically true
for any conformal field theory whose chiral algebra is generated
by a finite number of currents \Kir. Since all known unitary
RCFT's
have that property, this is probably a very mild
assumption.
Basic theorems on modular functions
can then be invoked to show that all coefficient functions must
be polynomials in $E_4$ and $E_6$.

We define the functions
${\cal E}_n$
as polynomials in $E_4$ and $E_6$ with total weight $n$.
These functions have one or more free parameters:
${\cal E}_{12k+l}$ depends on $k+1$
parameters for $l=0,4,6,8$ and $10$, and $k$ for $l=2$.

The characters out of which $P$ was built can be expanded in
traces over some fixed representation (called the {\it reference}
representation in the following).
Furthermore all traces
can be expressed in terms of a number (equal to the rank) of basic
traces $\Tr F^{s}$, where $s$
is equal to the order of
one of the fundamental Casimir operators
of the Lie algebra.
The reference representation must be chosen so that for all $s$
these basic traces are non-trivial and cannot be expressed in terms
of lower-order traces.
In the following all traces will be over the
reference representation unless a different one is explicitly
indicated.

Thus we arrive at the following expression for the character-valued
partition function
$$ P(q,F_1,\ldots,F_L)=e^{{1\over 48}E_2(q){\cal F}^2} (\eta(q))^{-8n}
  \sum_{m=0}^\infty \sum_i {\cal E}_{4n+m}(i) {\cal T}^m_i
    \ .\eqn\FirstPart $$
Here ${\cal T}^m_i$ denotes a trace of total order $m$, and $i$
labels the various combinations of traces of that order.

\subsection{Level zero}
Now we feed in some facts about the representations at the
zeroth level to determine some of the parameters
in the coefficient functions $\cal E$.
Since the ground state is a singlet representation of the theory,
it does not contribute to any of the higher traces.
This allows us to
rewrite   the partition
function in the following way
$$ \eqalign{
P(q,F) &= \exp\left({1\over  48} E_2(q) {\cal F}^2
           \right) \eta(q)^{-8n} \cr
&\times \Bigg\{ {\cal E}_{4n}(0) +  (E_4(q))^n
  \left[ \cosh\left({1\over48} \sqrt{E_4(q)} {\cal F}^2
\right) - 1\right]\cr
 &- ( E_4(q))^{n-3/2}  E_6(q)
  \left[ \sinh\left({1\over48} \sqrt{E_4(q)} {\cal F}^2\right)\right] \cr
 &+\sum_{m=2}^{\infty}\sum_i \Delta{\cal E}_{4n+m-12}(i)
  {\cal T}^m_i\ \Bigg\}  \ ,
 \cr  }\eqn\Traces$$
where ${\cal E}_{4n}(0)$ has  a leading term equal to 1.
The cosh and sinh terms, when expanded in $F$, produce
coefficient functions that are polynomials in $E_4$ and $E_6$ of
the correct weight. Their r\^ole is to cancel for the leading term
in $q$ the contribution of the exponential pre-factor.
We can take out a factor $\Delta=\eta^{24}$ from the remaining
coefficient functions, because we know that they must be proportional
to $q$. This leaves ${\cal E}_{4n+m}/\Delta$, which is an entire
modular function of weight $4n+m-12$ (since $\Delta$ has no zeroes).
Note that this shifts the weight of the unknown functions ${\cal E}$
by $-12$, effectively removing one free parameter for each coefficient
function.
The functions ${\cal E }_l$ exist only for $l=0$ and $l \geq 4$,
$l$ even.
For all other values that occur in the sum they must be
interpreted as 0.

\subsection{Level One }
Now consider the first excited level. Expanding \Traces\ to second
order in $F$ one gets
$$ {\cal N} + \left(15-{31\over6}n
+{{\cal N}\over 48} \right) {\cal F}^2
            + \sum_{\ell} \alpha_{\ell} \Tr_{\rm adj}
            F_{\ell}^2 \ .\eqn\Sone $$
Here $\alpha_{\ell}$ is the leading coefficient of
${\cal E}_{4n-10}(\ell)$ (times a factor for the conversion from
reference to adjoint representation).
This term vanishes if $n\leq 3$.
Since by construction the first excited level (the spin-1 currents)
consists entirely of adjoint representation of the Kac-Moody algebras,
the result should be equal to the Chern-character
$\Tr e^{F.\Lambda}$, where
$\Lambda$ is the adjoint representation matrix. Upon expansion this
yields, for non-Abelian algebras
$$ \sum_{\ell} \left( \dim_{\ell} + {1\over2} \Tr_{\rm adj}
F_{\ell}^2 \right) \ . \eqn\Stwo  $$
For $U(1)$ factors there is no
$F^2$ contribution in \Stwo, and any non-trivial representation can be
used for the other traces. Comparing \Sone\ and \Stwo\ we get,
for non-Abelian algebras
$$ \eqalign { \sum_{\ell} \dim_{\ell}  = \N~~~~{\rm and}~& \cr
 \left(30-{31\over3}n+{\N\over 24}\right) {k_{\ell} \over g_{\ell}}
                + \alpha_{\ell} &= 1 \cr }\eqn\TwoFormulas$$
For $n>3$ (\ie\ $c\geq 32$) the second equation simply determines
the coefficients $\alpha_{\ell}$, and one does not learn anything
about the possible Kac-Moody algebras. However, for $n\leq3$ these
coefficients are absent, and we get
$$ {g_{\ell}\over k_{\ell}} = 30-{31\over3} n + {\N\over  24} \ ,
\eqn\Rone $$ which is
independent of $\ell$. For $U(1)$ factors the right-hand side of
the second equation in \TwoFormulas\
is zero instead of one,
and $k_{\ell}/g_{\ell}$ is replaced
by the non-vanishing normalization constant $N_{\ell}$.
Hence in this
case we find (if $n\leq 3$)
$$\N=248n-720 \ .\eqn\Rtwo $$
This makes sense only if $n=3$. Then one finds that $\N=24$, and
substituting this into \Rone\ we conclude that any non-Abelian
factor that might still be present must have vanishing dual
Coxeter number. Since this
is not possible, all 24 spin-1
currents must generate $U(1)$'s. This saturates the central charge,
and hence
the entire theory can be written in terms of free bosons with
momenta on a Niemeier lattice. The only such lattice
with 24 spin-1 currents is the Leech lattice. Therefore this is the
only meromorphic $c=24$ theory in which Abelian factors appear.

Hence we may ignore $U(1)$'s from here on, and focus on non-Abelian
factors.
It is instructive to compute the total Kac-Moody central charge:
$$\eqalign{ c_{\rm tot} &= \sum_{\ell} {k_{\ell}  \dim_{\ell} \over
                              k_{\ell} + g_{\ell} }   \cr
     &= 24\ { {\cal N} \over 248 (3-n) + {\cal N}}\ , \cr } $$
which is valid  only if $n\leq 3$.
For $n=3$ we see that the result is always equal to 24,
which implies that
the Kac-Moody system ``covers'' the entire theory, and that the
unknown part of the theory defined above is necessarily trivial.
Our results so far
can be summarized as follows

{\it Let ${\cal C}$ be a modular invariant
meromorphic $c=24$ theory whose chiral algebra is finitely generated
and contains $\N$
spin-1 currents, with ${\cal N}\not=0$. Then either $\N=24$, and
${\cal C}$ is the conformal field theory of the Leech lattice, or
$\N>24$, and the spin-1 currents form a Kac-Moody algebra with
total central charge $24$. The values of $g/k$ for each simple factor
of this algebra are equal to one another,
and given by $\N/24-1$. }

For the special case of
simply laced, level-1 Kac-Moody algebras
(yielding even self-dual lattices) this result has been proved by
Venkov \Ven, who also observed that all the solutions to these
conditions correspond precisely to the Niemeier lattices.
Interestingly, Niemeier was able to classify all lattices without
knowing this fact.

This is all that can be learned from the trace identities at the
first level. The identities for higher-order traces involve
always unknown coefficients analogous to $\alpha_{\ell}$ above.
These coefficients can be determined and then used to
compute traces over the second excitation level.
\subsection{Level two}
At the second level we do not know in advance which representations
will appear, but at least we know which representations are
allowed to appear, namely all combinations of Kac-Moody representations
with total spin 2. For all types of traces of total
order 0, 2, 4, 6, 8, 10 and
14 we can compute the total value of that trace. This must be matched
by some combination of the spin-2 fields. By allowing arbitrary
positive integer coefficients for the multiplicity of each spin-2
field we get thus a set op equations for those multiplicities (of
course descendants of the spin-0 and spin-1 states must be taken
into account as well).

To write down these higher-order trace identities we first
need some definitions. The indices $J_{m_1,\ldots,m_r}(R)$ of a
representation $R$ of a simple Lie algebra are defined as
$$ \Tr_{R} F^m = \sum J_{m_1,\ldots,m_r} (R)
\prod_{i=1}^r \Tr (F^{s_i})^{m_i}\ , $$
where the traces on the right-hand side are over the reference
representation, and $\sum_i m_i s_i=m$. Here $r$ is the rank of
the Lie algebra, and the sum is over all combinations of basic traces
with the correct total order $m$. Note that with this definition
the indices depend on the reference representation.
For our purposes
it will be sufficient to consider the coefficients
$J_{m,0,\ldots,0}$, \ie\ the coefficient of $(\Tr F^2)^{m}$.
In a tensor product of $L$ Kac-Moody algebras we will
denote the coefficient of $(\Tr (F_1)^2)^{n_1}\times \ldots\times
(\Tr (F_L)^2)^{n_L}$ for a representation $R=(R_1,\ldots,R_{\ell})$
as $K_R(n_1,\ldots,n_L)$. Thus
$$ K_R(n_1,\ldots,n_L)=\prod_{\ell=1}^{L}
J_{n_{\ell},0,\ldots,0}(R_{\ell}) \ . $$

The second-level
trace identities can now be derived from \Traces. After a rather
lengthy computation we get
$$\eqalign{
\sum_R K_R(n_1,\ldots,n_L)&=\left[ \prod_{{\ell}=1,n_{\ell}\not=0}^L
{(2n_{\ell}-1)! \over 2^{ n_{\ell}-1} (n_{\ell}-1)!}
\left({k_{\ell} \over 2 N_{\ell} }\right)^{n_{\ell}}\right] \cr
&\times\left[ C_P - \sum_{{\ell}=1}^L \sum_{k=1}^{n_{\ell}}
{ 2^{k+1} n_{\ell} ! \over (n_{\ell}-k)!(P+k-1)! B_{2k}}
\left({2N_{\ell} \over k_{\ell}}\right)^k C_{k,\ell}
\right]      \ , \cr}     \eqn\HigherTrace
$$
which is valid if the total order, $P=\sum_{\ell} n_{\ell}$, is smaller
than or equal to 5.
The identity is valid for any
(non-trivial) choice of reference representation. The dependence
on this choice enters via the exponential ``anomaly'' factor in
\Traces, and manifests itself through the normalization constants
$N_{\ell}$. They are defined by the quadratic
trace of the reference representation matrices $\Lambda_{\ell}$
in the ${\ell}^{\rm th}$ group
$$ \Tr \Lambda^a_{\ell} \Lambda^b_{\ell} =  2N_{\ell} \delta^{ab}\ . $$
If one chooses the adjoint representation one must set
$N_{\ell}=g_{\ell}$ (the adjoint is a valid choice as long as only
quadratic traces appear). The coefficients $C_{k,\ell}$ are the
indices of the adjoint representation in the ${\ell}^{\rm th}$ factor,
\eg\ $C_{k,1}=K_{\rm adj}(k,0,\ldots,0)$, with respect to the
reference representation.
The coefficients $C_L$ in \HigherTrace\ are respectively equal to
196884, 32760, 5040, 720, 96, and 12 for $P=0,1,2,3,4$, and 5, where
$P$ is the total order of the trace, $P=\sum_{\ell} n_{\ell}$.
Finally, $B_{2k}$
are the Bernoulli numbers. There is an additional identity for traces
of order 14, which is a bit more complicated because the unknown
parameters of the $12^{\rm th}$ order traces must be cancelled.

There are many other higher trace
identities one could write down.
Unfortunately traces of higher
than second order are decomposable for the algebra
$SU(2)$, so that in that case
only combinations of second order traces can be used. Most of the
accidental solutions to the first level trace identities are precisely
due to $SU(2)$'s. For this reason it was not worthwhile to consider
higher trace identities.

\subsection{Solution methods}
Although all solutions to the first level identities correspond
to Niemeier lattices if one considers only simply laced algebras,
this is not true in general. Our first priority is therefore to
rule out as many of the 221 Kac-Moody combinations as possible. To do
so,
we would like to solve the second level
equations for each of the 221 Kac-Moody
combinations, or prove that no solution exists. Unfortunately,
the number of spin-2 fields is often
much too large. This problem can be solved by symmetrizing the
equations over identical factors in the tensor product, which usually
reduces the number of variables much more than the number of equations.
Obviously,
the existence of a solution to the symmetrized equation is a necessary,
but not a sufficient condition for the existence of a solution to the
full set of equations.
The number of variables is typically about 50, and at most 288.
The number of equations is usually larger than the number of variables,
although not all {\it a priori} distinct equations are independent. For
$\N < 36 $ one finds, however, often fewer  equations than variables.
One of the worst cases has 63 equations for 248 variables.

There are several ways of dealing with these equations. The simplest
procedure is to compute the greatest common divisor of the coefficients.
Consider for example the combination $(B_{4,1})^4 A_{6,1}$ with
$\N=192$.
The right hand side of the
zeroth order trace identity is computed by subtracting the
descendant contributions from 196884, which yields:
$$\eqalign {196884 - &(4\times 36 + 48 ) \cr
                   - &(4\times 36 \times 48 + 6 \times 36^2 ) \cr
                   - &(4\times (\half[36\times 37]-495)
                       +       (\half[48\times 49]-735)) = 180879 \cr}$$
In this case there is just one spin-2 field available, as one
may easily check, namely a combination of the four vector
representations
of $B_{4,1}$. This has a ground state dimension $9^4 = 6561 $, which
is not a divisor of 180879. Hence for this combination there is no
solution to the trace identity, and therefore no conformal field
theory can exist. There are three other $\N=192$ combinations,
for which solutions {\it do} exist. They correspond to the
Niemeier lattice $(A_{6,1})^4$, the $\Zbf_2$-twisted Niemeier lattice
$(B_{4,1})^2D_{8,2}$ (derived from $D_{9,1}A_{15,1}$)
and a new theory $B_{4,1}(C_{6,1})^2$.

Many combinations with large values of $\N$ can be ruled out by
this sort of argument. It occurs rather frequently that
for one or more kinds of traces all allowed
spin-2 fields have a common factor, which does not divide
the right hand side. This is the easiest way to rule
out `fake' solutions (\ie\ solutions to the level-one conditions
without a corresponding modular invariant partition function).

If this common divisor method does not yield inconsistencies, ruling
out fake solutions becomes more difficult. Very often the following
method works. We know that all unknowns must be positive integers.
Furthermore they are bounded from above since the total number of
states at the second level must be 196884.
Using a set of bounds on all variables, one can compute new
bounds from the equations, or from suitably chosen linear combinations
of the equations.
Very often one ends up with
an inconsistency for the bounds on some variable.

If this does not work, one can make use of the knowledge that the
unknowns should be integers. To do so,
one can use standard methods for solving
linear equations.
If the number of
variables is larger than the number of equations,
the best one can do is try all allowed integer values (within the
boundaries) for the remaining variables. This might easily have
failed, since the number of possibilities grows exponentially with
the number of undetermined variables, but luckily in the few cases
where it was necessary, it was possible. The  final result is
that only 69 of the 221 combinations remain.

Of course all these computations were done with a computer. This
has two disadvantages. First of all, it is not possible to
present details of the elimination process as we did for the
example above. Thus there is no ``presentable'' proof.
Secondly, one has to worry about programming errors and accuracy.
Most errors of the former kind would almost certainly affect one
of the known solutions, and therefore such errors are not very
likely. Accuracy becomes an important
issue especially in those cases where it was necessary to
solve the
linear equations.
The computations were done in FORTRAN using
extended precision floating point arithmetic.
Integer arithmetic is exact, but the integers
become rapidly extremely large, and easily exceed the maximum value
(i.e. $\approx  10^9$) with disastrous consequences. The main worry
with floating point arithmetic is loss of accuracy. However, the
32-digit accuracy that was used should be more than sufficient.
Nevertheless, it might be worthwhile
to repeat the computations with
an algebraic program, to solve the linear equations exactly.

\subsection{Modular Invariant Partition Functions}

For the remaining 69 combinations we expect a conformal field theory
to exist, since that is the only way to make sense of the fact that many
equations can be satisfied with simple integers. Therefore we expect
that there must exist a modular invariant partition function. The level-2
solution gives us partial information about that function, but we still
have to ``unsymmetrize'' it (if there are several identical factors) and
determine the higher spin content. This task seems hopeless at first, but
the problem is simplified drastically by simple currents. If among the
known spin-2
fields that appear there are one or more simple currents, then
we know that

\item{\bullet}
Fields with fractional charges with
respect to those simple currents cannot
appear (this simply follows from the requirement op locality of
the operator algebra).

\item{\bullet}
The multiplicity of all fields is constant on the simple current orbits.
This can be proved using the form of
the matrix $S$ due to simple currents.

Each of these two points allows us to reduce the number of primary fields
that we need to consider by a factor $N$, where $N$ is the order of the
simple current
(the reduction is slightly less when there are fixed points).

Here again a bit of luck was needed to
make the problem manageable. In some
cases, the spin-2 fields do not include any simple current. Fortunately
it was possible to investigate those without any reduction in the
number of fields. In other cases the reduction of the number of fields
was large enough to make the problem amenable to computer calculation.
After taking into account all known simple currents, the effective
number of integer spin fields was less than 250, except for one case
($(A_{1,4})^{12}$) with 1147 integer spin fields, which required
special treatment.
In these calculations issues related to computer accuracy are far less
important, since problems of this kind are far more likely to
eliminate valid solutions than to generate invalid ones.

For each of the 69 remaining cases we found precisely one meromorphic
modular invariant (a few of
the Niemeier lattices were not investigated, because they have in any
case already been classified completely). The complete list appears
in \SchM, and will not be repeated here.
Thus, if the monster module is indeed unique,
and if there is precisely one MCFT per modular
invariant, then the total number of such theories is 71.

Some of the Kac-Moody combinations that were
ruled out by the level-two trace identities had a small enough
number of integer spin fields to be checked explicitly for meromorphic
modular invariants. Still others could be checked under the
additional assumption that some integer spin simple currents appear
in the chiral algebra. Several meromorphic
invariants were indeed found, but they all involve spin-1 currents.
This means that they have a larger Kac-Moody sub-algebra, and are
embedded conformally in some other meromorphic theory. For example,
all theories with ${\cal N}=48$ can be embedded in $D_{24}$ (using
the embedding $SO(\dim(H)) \supset H$, with  the vector representation
branching to the adjoint of $H$). These embeddings do not appear
as a solution to the trace identities, since they violate the
assumption that the spin-1 currents are all absorbed into adjoint
representations. The fact that they are found as modular invariant
partition functions whenever expected, and that no other invariants
are found is an important check on the calculations. Unfortunately
this check is not available in all cases, because the number of
integer spin currents is simply too large.

\subsection{Construction}

Explicit constructions exist for 39 of the 71 theories, and most likely
for two additional ones.
The 24 lattice theories and the 15 $\Zbf_2$-twisted
theories have been constructed \DGM.
The $E_{8,2} \times B_{8,1}$ theory
can be obtained from a (different)
$\Zbf_2$-orbifold twist \DiH\ or can be
built out of free fermions \KLTE. Since  free fermion theories (even with
real boundary conditions) can be formulated on arbitrary genus
Riemann surfaces, there should be no difficulty in writing down
a multi-loop partition function (see \multref\ABK{\KLST\BDGA}).
By factorization, that implies the
existence of all correlation functions on arbitrary Riemann surfaces,
which is tantamount to existence of the theory.

An investigation of other orbifold twists of meromorphic CFT's was
presented by P. Montague \Mon. Unfortunately, the consistency conditions
for such orbifold are apparently difficult to verify. Many candidate
theories appear that do not correspond to an allowed Kac-Moody
combination, and that therefore must be inconsistent.
In \Mon\ four theories are found that appear on the list of 221
combinations satisfying the level-1 conditions, but only one of
those ($E_{6,2}A_{5,1}C_{5,1}$) survives the level-2 conditions. The
latter is obtained by starting with the Niemeier lattice
$(E_{6,1})^4$, with a $\Zbf_2$ twist consisting of an interchange of
two $E_6$ factors and two different involutions on the other  two
$E_6$'s. If the consistency of this procedure can be proved, we would
have a construction of one more theory on the list.
(The other three candidates were all obtained by
applying an extra twist to some of the 15 theories of \DGM.)

This still leaves 30 theories to be constructed. Perhaps they can
all be obtained using orbifold twists on some other theory, but
there is at present no evidence to support this. It might be
worthwhile to explore how many of them can be constructed using
free fermions. This set probably includes
the Niemeier lattices (most of them can indeed be constructed out of
free fermions with complex boundary conditions, but I have not checked
this for all of them),
then most likely also the 15 twisted lattice theories,
as well as the $E_{8,2} B_{8,1}$ theory. Even though
combinations like $F_{4,6}\times A_{2,2}$ consist of factors that
cannot separately be constructed out of free fermions (the
central charges are ${104\over5}$ and ${16\over5}$), perhaps the
combination {\it can} be obtained in this way.

It is
possible to apply the methods of \KLTA\ABK\ to construct
systematically {\it all} possible meromorphic free fermionic theories
with $c=24$. However, the still preliminary results are rather
disappointing. So far all allowed fermion boundary conditions have
been constructed, but not all the allowed phases have taken into account.
For complex boundary conditions I do seem to get all
the Niemeier lattices (for example the not straightforwardly
fermionic $A_{24}$ theory has appeared). However,
there are relatively few choices  of boundary
conditions for real, unpaired fermions. Among the resulting spectra
I do find a surprisingly simple realization of the monster module (which
is probably already known to mathematicians, although the precise
correspondence is hard to establish), as well as the $E_{8,2} \times
B_{8,1}$ theory and some twisted Niemeier lattices. Nothing genuinely
new has emerged so far, however. Some more details may appear in a
future publiication.

\subsection{The number 71}

One may hope that the complete list of 71 theories displays some
interesting underlying structure, just as the list of $SU(2)$
modular invariants revealed a relation to ADE-Dynkin diagrams. Indeed,
the list of 24 even self-dual lattices might make more sense when it
is embedded in the list of 71 meromorphic
CFT's. Clearly what is missing is some organizing principle that
makes our result something more that an uncorrelated list of
modular invariants.

So far not much has emerged. There are, however, two
rather intriguing, though highly speculative observations
concerning the total number
of (candidate) MCFT's, 71. First of all
this number is equal to the largest prime factor in the
number of elements of the monster group,
$$ 2^{46}\cdot 3^{30}\cdot 5^{9}\cdot  7^{6} \cdot 11^3 \cdot 13^3
\cdot 17 \cdot 19 \cdot 23 \cdot 31 \cdot41 \cdot 47 \cdot 59
  \cdot 71 \ . $$
It is totally mysterious what this might mean, if anything.

A second observation concerns the equation
$$ \sum_{i=0}^N i^2 = M^2 \ , $$
with $N$ and $M$ positive integers. The only solutions I know to
this equation are two trivial ones,
$(N,M)=(0,0); (1,1)$, and a unique non-trivial solution,
$(24,70)$.
The latter solution does indeed play a r\^ole in this context:
the Leech lattice can be characterized as a subset of the lattice
$\Gamma_{25,1}$ (which can be described as
$D_{26}$, but with a Lorentzian signature, and with a
spinor conjugacy class added), consisting of a set of points
orthogonal to a lightlike
vector $\lambda$ modulo that vector. One choice
for $\lambda$ is the vector $(0,1,2,3,\ldots,24;70)$.
This suggests that there might exists some `natural' correspondence
between the integers from 1 to 71 and the 71 MCFT's, with
70 assigned to the Leech lattice and perhaps
71 to the monster module.
The problem with this idea is that the light-like vectors specifying the
Niemeier  lattices within $\Gamma_{25,1}$ are by no means unique.
One would have to find some natural choice among the infinity
of possibilities. Furthermore one has to find a
CFT-generalization of
the concept of a light-like vector and a lattice `orthogonal'
to it.

Clearly, if any of these observations is more than just a
numerological coincidence, this would be an extremely exciting
development.
\par \penalty-4000\vskip\chapterskip
   \spacecheck\referenceminspace \immediate\closeout\referencewrite
   \referenceopenfalse
   \line{\fourteenrm\hfil REFERENCES\hfil}\vskip\headskip
   \endlinechar=-1
   \input referenc.texauxil
   \endlinechar=13
   
\end